\documentclass[aps,twocolumn,preprintnumbers,%
superscriptaddress,floatfix]{revtex4}

\usepackage[dvips]{graphics}
\usepackage{amsmath,amsfonts,bm}
\usepackage{epsfig}

\begin{document}
\newcommand{\be}{\begin{eqnarray}}
\newcommand{\ee}{\end{eqnarray}}
\def\lsim{\mathrel{\rlap{\lower3pt\hbox{\hskip1pt$\sim$}}
     \raise1pt\hbox{$<$}}} 
\def\gsim{\mathrel{\rlap{\lower3pt\hbox{\hskip1pt$\sim$}}
     \raise1pt\hbox{$>$}}} 
\def\N{${\cal N}\,\,$}
\def\bi{\bibitem}\def\la{\langle}\def\ra{\rangle}
\newcommand\<{\langle}
\renewcommand\>{\rangle}
\renewcommand\d{\partial}
\newcommand\LambdaQCD{\Lambda_{\textrm{QCD}}}
\newcommand\tr{\mathrm{Tr}\,}
\newcommand\+{\dagger}
\newcommand\g{g_5}

\title{NA60 and BR Scaling In Terms of The Vector Manifestation: Formal Consideration}

\author {G.E. Brown}
\affiliation { Department of Physics and Astronomy\\ State
University of New York, Stony Brook, NY 11794-3800}
\author{Mannque Rho}
\affiliation{ Service de Physique Th\'eorique,
 CEA Saclay, 91191 Gif-sur-Yvette c\'edex, France}
\begin{abstract}
The arguments developed in the preceding article on how BR scaling
would predict for dilepton production in heavy-ion collisions,
e.g., NA60, are augmented with more precise and rigorous
arguments.
\end{abstract}


\newcommand\sect[1]{\emph{#1}---}

\maketitle

In the preceding article~\cite{BR:na60-b}, we gave an extremely
simplified model argument why the green curve erroneously
attributed to BR scaling in the NA60 results~\cite{na60} has
nothing to do with what we consider to be BR scaling proposed in
1991~\cite{BR91} and modernized recently with hidden local
symmetry theory of Harada and Yamawaki~\cite{HY} with vector
manifestation~\cite{HY:VM}. In order to make our arguments
accessible to non-theorists, we used in \cite{BR:na60-b} arguments
that are somewhat over-simplified and hence lacking rigor and
precision. In this paper, we supply the missing details with more
precise definitions and reasoning, indicating which arguments are
rigorous and which are not and what needs to be done to make them
firmer. The conclusions presented in \cite{BR:na60-b} remain
qualitatively unchanged although some were heuristic and
incomplete, so the readers who are convinced by the arguments
given there could skip reading this paper.

In our approach to the physics of matter under extreme conditions,
the correct implementation of the basic premise of BR scaling is
mandatory not only to the immediate problem of theoretically
interpreting the NA60 data or other dilepton data but also to the
whole gamut of RHIC physics since the vector manifestation (VM in
short, see below) of chiral symmetry discovered in hidden local
symmetry theory by Harada and Yamawaki~\cite{HY:VM}, we believe,
is relevant not only in the hadronic phase into which the high
temperature and high density matter produced in relativistic
heavy-ion collisions or superdense matter formed in gravitational
collapse in compact statrs evolves but also for understanding the
structure of the ``new form" of matter found in the chirally
restored phase as discussed in a series of recent papers by Brown
et al. addressing RHIC physics~\cite{BGR,BLR05,BLR-star}. In this
paper we will focus on aspects related to dilepton production.

There are three key elements of hidden local symmetry theory with
vector manifestation (HLS$^{VM}$ in short, to be distinguished
from a more general HLS theory described below) relevant to
dilepton production in heavy-ion collisions: (1) vector dominance
(VD) in the photon coupling to matter, (2) the distinction and
role of the ``parametric" mass and the physical (or pole) mass of
the vector meson and (3) the ``sobar" excitations and ``fusing" of
BR scaling and sobar configurations. All three are essential in
describing dilepton production but have not been properly taken
into account in most of the works available in the literature.

$\bullet$ Vector dominance (VD).

The principal point of vector dominance in the process at issue is
that vector dominance is mostly violated when the photon couples
to matter in medium. For this we need to understand when VD is
operative and when it is not. Let us explain how this issue comes
about.

\begin{figure}[htb]
\centerline{\epsfig{file=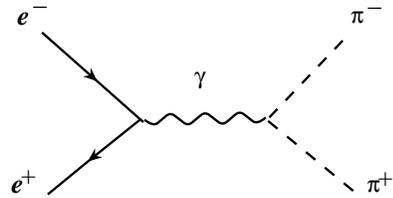,width=10.cm,angle=0}} \
 \vskip -4.0cm
 \caption{Direct photon coupling to the pions which is absent
in the holographic dual QCD theory and also in Harada-Yamawaki's
HLS theory for $a=2$.} \label{photon}
 \end{figure}

In a recent important development in holographic dual
QCD~\cite{dualQCD} (an approach to QCD emerging from AdS/CFT
duality in string theory), low-energy QCD is given by a multiplet
of (pseudo)Goldtone bosons and an infinite tower of massive vector
bosons with local gauge invariant coupling. The infinite tower
results from the 5-D Yang-Mills theory given by the dual bulk
sector when the 5-D nonabelian gauge theory is compactified \`a la
Kaluza-Klein to 4-D gauge invariant theory. This is a generalized
hidden local symmetry theory in the sense that an infinite tower
of gauge fields are involved. Let us call this theory HLS$^{AdS}$.
It turns out that the 4-D theory is entirely vector-dominated.
What is significant for our case is that the photon coupling to
hadrons is $completely$ vector-dominated. That means that the
direct photon coupling to charged pions as depicted in
Fig.\ref{photon} is simply absent in the theory. However when
truncated at a finite number of vector mesons, say, $\rho$,
$\omega$, $\phi$, it is natural to expect that vector dominance is
lost in general. However surprisingly while it is lost in certain
processes (e.g., the EM form factor of the nucleon to be discussed
below), it is preserved in some others (e.g., the
$\omega\rightarrow 3\pi$ decay and some anomalous photon processes
mediated by the Wess-Zumino term). In Harada-Yamawaki theory
HLS$^{VM}$, one makes the truncation at the lowest members of the
tower, $\rho$, $\omega$ and $\phi$ and integrates out all the rest
of the infinite tower at some scale $\Lambda_M$ and makes the
Wilsonian matching to QCD at $\Lambda_M$ for an ultraviolet
completion. In such a theory the vector dominance is of course
necessarily incomplete or violated. Now the theory has three key
constants (apart from quark masses etc), the gauge coupling $g$,
$F_\pi$ which is connected to the pion decay constant $f_\pi$ and
$a$ which is the ratio $(F_\sigma/F_\pi)^2$ -- where $F_\sigma$ is
the decay constant of the would-be Goldstone boson that makes up
the longitudinal component of the massive vector meson. The
renormalization group equation (RGE) trajectories for the
constants in this HLS$^{VM}$ theory are in general quite complex.
However when constrained to QCD by matching of the correlators,
things become very simple: the RGE for the gauge coupling $g$
flows to the fixed point $g=0$ and that for the constant $a$
flows to 1 whereas unconstrained, they can flow to a variety of
fixed points. It is the constraint imposed by matching with QCD
that leads uniquely to the fixed points $(g^*, a^*)= (0, 1)$. As
for $F_\pi$, the constraint with QCD does not lead to any special
value in temperature (while it goes to zero in density). However
the physical pion decay constant $f_\pi$ related to $F_\pi$ with
quadratically divergent loop corrections does go to zero at the
fixed point $(g^*, a^*)= (0, 1)$. This fixed point to which
hadronic system is driven when chiral restoration is reached is
called ``vector manifestation fixed point."

We do not know whether -- and how -- one can probe what
corresponds to the VM fixed point in HLS$^{AdS}$. That requires
higher order computations in the bulk sector which nobody knows
how to do. We shall therefore consider Harada-Yamawaki HLS$^{VM}$
theory. For this, let us model what happens using only the
low-mass mesons $\pi$, $\rho$ etc  in phenomenology which should
be known to everyone. In HLS$^{VM}$, the photon coupling to
hadrons is given by
 \be
\delta {\cal L}=-2eagF_\pi^2 A^\mu{\rm Tr} [\rho_\mu Q] +2ie
(1-a/2)A^\mu {\rm Tr}[J_\mu Q]\label{current}
 \ee
where $A_\mu$ is the photon field, $\rho_\mu$ stands for the
vector fields $\rho$, $\omega$ and $\phi$, $J_\mu$ is the vector
current made up of the chiral field, i.e., pions and $Q$ is the
charge matrix $Q=\frac 13 {\rm diag} (2 -1 -1)$. Now the vector
dominance is achieved in this theory when $a=2$ for which the
second term in (\ref{current}) which corresponds to
Fig.\ref{photon} vanishes. However as shown by Harada and
Yamawaki~\cite{HY}, $a=2$ is {\it not} on a stable trajectory of
the RGE for $a$ and that nature is realized by vector dominance in
pionic processes is merely an ``accident" rather than required by
QCD. In fact generic hadronic systems tend to quickly go to $a=1$
under normal conditions as discussed in \cite{DSB-mr} which
implies in particular a 50/50 electromagnetic
coupling~\cite{BRPR1,BRDD} of hadrons directly and through the
vector meson to leptons. This phenomenon is known to take place in
even free space. For instance, the nucleon form factor is dipole
in nature, not monopole as would be given by vector dominance.
This and other phenomena discussed in \cite{BRPR1,DSB-mr} show the
vector dominance used at zero density and temperature to be very
fragile and easily violated in the presence of matter and
temperature. Indeed Harada and Sasaki have shown that in a heat
bath, vector dominance is maximally violated with the constant $a$
going to 1 ~\cite{harada-sasaki}. This is not surprising since
vector dominance does not appear on the RG required by QCD.

In heavy-ion processes, temperature and density are intricately
correlated, so it is difficult to make a precise analysis of how
$a$ evolves as a function of $both$ temperature and density. But
we expect that the combination of the two will speed up the flow
of $a$ to 1 compared with the influence of either alone. What this
means specifically in dilepton processes is that the vector
dominance part of the Lagrangian has coefficient 1/2 as compared
with the one used by most workers so that the dilepton production
should be 1/4 of that calculated using vector dominance. The part
of the Lagrangian in which the hadrons couple directly to
dileptons will provide a background of the $\rho$-mesons, because
the latter is not involved in the interaction.

The conclusion that we can draw from the above argument is that
the overall use of vector dominance which has been the common
practice in the field would lead to a curve in dilepton production
as much as 4 times higher than would be given by the Lagrangian
(\ref{current}) that is predicted by Harada-Yamawaki HLS$^{VM}$
theory.

$\bullet$ Intrinsic background dependence (IBD) and parametric
mass.

In HLS$^{VM}$ theory, the pole mass of the vector meson that
appears in the current-current correlation functions measured by
the experiment is given by two terms: one the ``bare" or
``parametric" mass and the other the thermal (and/or dense)
loop-correction term, all of which are temperature/density
dependent,
 \be
m_V = m_{bare} +\Delta M\label{mass}
 \ee
with
 \be
m_{bare}&=& \sqrt{a}gF_\pi.
 \ee
Now the first term $m_{bare}$ consists of the parameters that
figure in the bare Lagrangian which is determined in HLS$^{VM}$ at
the matching scale $\Lambda_M$ whereas the second term is a loop
correction computed with the same parameters. Since the
correlators are Wilsonian-matched between the effective field
theory (EFT) sector and QCD sector, the parameters in the EFT
sector can be expressed in terms of the quantities that appear in
the QCD sector, namely color-gauge coupling, quark and gluon
condensates. If the matching is done in medium, then the
condensates will inevitably depend on the background that defines
the medium, namely, density, temperature etc. This means that the
parameters $a$, $g$ and $F_\pi$ in the EFT sector will depend on
the same background. This is referred to as ``intrinsic background
dependence (IBD)." When we say ``BR scaling mass," we mean
$m_{bare}$, not $m_V$ (except near the critical point). It is this
IBD term that is directly locked to the chiral symmetry property
of the vacuum. Near the critical point, HLS$^{VM}$ says that both
$m_{bare}$ and $m_V$ go proportional to the quark condensate
$\la\bar{q}q\ra$ and hence go to zero as BR scaling does.

If one were to do a naive tree-order calculation and ignore loop
corrections, only the first term would contribute. If this were
the entire story (or dominant story), the physical vector meson
mass would then be following the quark condensate. However this is
certainly wrong in hot matter as pointed out very clearly by
Harada and Sasaki~\cite{harada-sasaki} and also near chiral
restoration in dense matter as pointed out by Harada et
al~\cite{HKR}. Let us focus here on the temperature effect. We
will return to the density case later. In a heat bath even at low
temperature, the (second) loop corrections are mandatory for
consistency with the symmetry of QCD. In fact it is in combination
of the two terms that the pole mass of the vector meson
$increases$ $\propto T^4$ near zero temperature with no $T^2$ term
present as required by low-energy theorem~\cite{DEI}. Thus on the
one hand, the IBD is required by the matching of the correlators
at $T_C$ to QCD~\cite{harada-sasaki} and on the other hand, the
IBD is {\it not} what is measured directly in experiments. As
$T_c$ is approached, both the bare mass term and the loop
corrections go to zero $\propto \la \bar{q}q\ra\rightarrow 0$. In
this case the pole mass does directly reflect on chiral structure
as does BR scaling. {\it Only in the vicinity of $T_c$ does BR
scaling manifest itself transparently in the pole mass of the
vector meson in a heat bath.}

Now what do we know about the temperature dependence of $m_{bare}$
and $m_V$? While it is clear that the vector meson pole mass will
not simply follow the order parameter of chiral symmetry as the
$m_{bare}$ does in HLS$^{VM}$, we have no theoretical information
as to what happens away from $T=0$ and $T=T_c$. Nobody has
calculated it yet although it is a doable calculation. Fortunately
lattice calculations can provide the necessary information. As
described in \cite{BLR05}, we learn from Miller's lattice
calculation of the gluon condensate~\cite{miller} that the soft
glue starts to melt at $T\approx 125$ MeV. The melting of the soft
glue, which breaks scale invariance as well as chiral invariance
$dynamically$ -- and is responsible for the hadron mass -- is
completed by $T_c$ at which the particles have gone massless. The
gluon condensate which remains unmelted above $T\sim T_c$
represents the hard glue, or what is called ``epoxy," which breaks
scale invariance $explicitly$ but has no effect on the
(dynamically generated) hadron mass. Coming down in temperature in
heavy ion processes, $T\approx 125$ MeV is therefore the ``flash"
temperature at which the vector meson recovers 95\% of its
free-space mass. We see that the melting of the soft glue is
roughly linear, implying that the meson masses increase linearly
from near zero at $T_c$ to the flash temperature. This then
suggests that going up in temperature, nothing much happens to
$m_V$ until $T\sim 125$ MeV. It is this scenario that provides an
extremely simple explanation~\cite{BLR-star} for the observed STAR
$\rho^0/\pi^-$ ratio and HBT~\cite{star}.

In fact, an approximate calculation~\cite{masa} of the thermal
loop terms based on the work of Harada and
Shibata~\cite{harada-shibata} finds that up to the flash
temperature of $T\sim 125$ MeV, the $\Delta M$ term in
(\ref{mass}) is positive and small in magnitude, $\sim +10$ MeV.
Since the bare mass $m_{bare}$ is flat up to that temperature as
indicated from the lattice results, it is very reasonable to
assume that the vector meson pole mass does not change appreciably
up to the flash point.

Let us consider now the density effect. The density probed in the
dilepton experiments is in the vicinity of nuclear matter density
so it is not high. Now in the bare mass term, we expect that
$\sqrt{a}g$ changes little up to nuclear matter
density~\cite{BRDD}, so
 \be
\Phi\equiv \frac{m^*_{bare}}{m_{bare}}\approx
\frac{f_\pi^*}{f_\pi}
 \ee
At $n\approx n_0$, we know that this should be $\sim
0.8$~\cite{friman-rho} and expect that slightly above $n_0$ which
is reached by the dilepton processes, the linear drop $\Phi
(n)\approx 1-0.2(n/n_0)$ would be reasonable. The dense loop term
$\Delta M$ is also positive and small as in the case of
temperature in the regime of the density involved and it will go
to zero at the VM fixed point. This means that the mass drop in
density is expected to be $less$ than 20\% at normal nuclear
matter density. This is indeed what one finds at $T\sim 0$ in the
CBELSA/TAPS experiment~\cite{omega} on the $\omega$ meson as well
as the KEK experiment~\cite{KEK} on $\rho$ and $\omega$.

The conclusion is that combining the temperature effect and
density effect, one expects the mass shift in the vector mesons in
the NA60 experiment to be small, if any. {\it This does not mean
that BR scaling is absent in the process.}

$\bullet$ Fusion with ``sobar" configurations.

In the HLS$^{AdS}$ theory given by holographic dual QCD,
low-energy physics is entirely described by a multiplet of
pseudo-Goldstone pseudoscalars, an infinite tower of massless
vector bosons and a multiplet of Goldtone scalar bosons that are
to be higgsed. In unitary gauge, the vector mesons become massive
with the scalars eaten up. Since this represents the entire story
of QCD at low energy with no fermions in the picture, we must
conclude that baryons $must$ emerge as skyrmions. This has been
pointed out by several theorists~\cite{dualQCD,hill,son}. It is
clear what one should do: one should calculate spectral functions
measured in experiments either from this generalized hidden local
symmetry theory with an infinite tower of vector mesons or more
practically from the Harada-Yamawaki HLS$^{VM}$ of the lowest
vectors matched to QCD. The first serious effort was made in
\cite{sliding} along this line using the Skyrme model that
contains only the pion field and an initial attempt is being made
with HLS$^{VM}$ theory but we are still very far from a realistic
calculation that can be confronted with experiments. As mentioned,
if the temperature is near $T_c$ where the vector meson mass
becomes comparable to the pion mass, then Harada-Yamawaki HLS/VM
theory should be reliable enough. See \cite{kaon,DSB-mr} for more
discussions on this point. However near $T\sim 0$ or far away from
$T_c$, Harada-Yamawaki's HLS theory without fermion (baryon)
degrees of freedom cannot fully capture the physics of all
channels. For instance, it cannot describe meson condensations in
dense matter unless baryon degrees of freedom are incorporated.
This is because we cannot ignore certain low-energy particle-hole
excitations that have the same quantum numbers as the mesonic
degrees of freedom we are looking at. Notable examples are the
``sobar" excitations, e.g., the $\Delta$-hole excitations at about
300 MeV in the pion channel, the $N^* (1520)$-hole at about 580
MeV in the $\rho$ channel etc. As in condensed matter physics,
these excitations are expected to be very important at temperature
or density at which HLS/VM is not of dominant influence.
Incorporating these sobar configurations in studying spectral
functions is referred to as ``fusing"~\cite{BRPR2,BRDD}. How to
treat the fusing of sobar degrees of freedom with elementary
excitations was first discussed in \cite{sobar} using a field
theory technique with a toy model. If one were to generate baryons
$N$, $N^*$ etc from HLS theory as skyrmions and do an effective
field theory \`a la Harada and Yamawaki, then the intrinsic
density/temperature dependence that underlies BR scaling would be
well-defined and hence the fusing could be done in a consistent
way. This has not been done yet. Positing the presence of the
baryons and the Fermi surface as was done in \cite{BRPR2,BRDD} is
undoubtedly {\it ad hoc} but that's the best one can do at the
moment.

The above caveat notwithstanding, our assertion is that in
describing dilepton process such NA60, BR scaling reflecting on
chiral symmetry in hot matter that is probed -- which is a
parametric property of the effective Lagrangian -- should be
implemented in the fusing of the sobar and elementary
configurations. This phenomenon cannot be captured  {\it except
perhaps very near the critical point} by just a single elementary
field with a dropping mass as has been done so far. This aspect
has been amply emphasized in review articles \cite{BRPR2,BRDD}.

In short, it would be too naive to expect that the shape of the
spectral function measured in the NA60 provide a $direct$
information on the chiral structure of the hadronic system. A mere
shift of the peak either way cannot be taken as a signal for or
against BR scaling.

$\bullet$ ``Seeing" chiral restoration, partial or full.

There is an intense effort among both theorists and experimenters
to find a direct signal for both what is called ``partial" chiral
restoration and ``full" chiral restoration. The former is looked
for in what are considered to be ``physical" in-medium quantities
like pion decay constant, vector meson mass etc. and the latter in
probing matter under extreme conditions such as in heavy ion
collisions or in compact stars.

Needless to say, what one measures in experiments are correlation
functions, not such theoretical quantities like in-medium mass,
in-medium decay constant etc. How to interpret the behavior of
masses and coupling constants in one's theory depend on what
theory one is using. Now in connection with chiral symmetry one
would like to have a set of parameters that reflect in a {\it
known way} on the order parameter of the symmetry such as for
instance the quark condensate in the case of chiral symmetry in
QCD and to express in a consistent way the correlators one would
like to study. To learn how chiral symmetry manifests itself as
the conditions in the vacuum are changed, one would have to map
out the parameters that are locked in a known way to the order
parameter, i.e., quark condensate. BR scaling is one such
parametrization, not necessarily the only one. They are the ones
which will ultimately provide the desired information. It is not
necessarily physical variables themselves that will do so. The
theoretical situation is much clearer when one is near the phase
transition, as we stated, thanks to the vector manifestation in
the case of HLS$^{VM}$ theory. But it cannot be so away from the
transition point.

To illustrate that one can easily arrive at a completely wrong
picture if one ignores subtlety involved in the notion of hadron
effective mass -- whether it is BR scaling mass or Landau
effective mass, let us look at the well-known case of the EM
orbital current and the in-medium mass of a nucleon sitting on top
of the Fermi sea which was in the past cited as an evidence
against BR scaling. Consider a chiral Lagrangian in which BR
scaling is suitably implemented. A nucleon in nuclear matter
described as a quasi-particle with such a Lagrangian will carry a
BR scaling mass $m^*$ which drops as density increases in some
proportion to the quark condensate. Assuming as is often done in
nuclear physics that the impulse approximation is valid in
response to the slowly varying EM field, one may write the
isoscalar convection current for the nucleon with the momentum
$\mathbf{k}$ as~\cite{footnote2}
 \be
\mathbf{J}=e\frac{\mathbf{k}}{2m^*}.
 \ee
Now suppose that experimentalists measure the isoscalar orbital
gyromagnetic ratio $g^0_l$ defined by
 \be
\mathbf{J}=eg^0_l \mathbf{k}/m_N
 \ee
where $m_N$ is the free-space proton mass $m_N=938$ MeV. Their
experiment will of course yield (as we all know)
 \be
g^0_l=1/2
 \ee
and $not$ $g^0_l=\frac 12 (m_N/m*) > 1/2$, as the naive
calculation would give. They might then be tempted to say ``BR
scaling is ruled out by the experiment on the gyromagnetic ratio
in nuclei."

This conclusion is completely wrong. In Fermi liquid theory, one
gets the convection current given by the free-space mass due to a
well-known mechanism, variously attributed to charge conservation,
Galilei invariance etc. In condensed matter physics it is related
to Kohn theorem for cyclotron frequency of an electron. In
many-body theory language, it is the effect of ``back-flow."  In
the case with BR scaling in a chiral Lagrangian, it is the chiral
Ward identity that assures the correct answer as shown in
\cite{friman-rho}.

This example illustrates the danger in jumping to a conclusion
prematurely. There are other such cases. For example, a similar
misleading conclusion could have been arrived at in spin
observables in proton-nuclear scattering as discussed in
\cite{BRDD}.

$\bullet$ Concluding remarks.

We have discussed in a more precise language than in the preceding
article three important features that characterize BR scaling as
interpreted in terms of hidden local symmetry theory with vector
manifestation, with a focus on their role in describing the NA60
dilepton process. First, the parameter $a$ going to 1 in hot and
dense matter will maximally violate the vector dominance and hence
cut down the dilepton cross section. Secondly parameterizing the
temperature dependence of the pole mass of the vector meson
$m_V^*/m_V$ as $(1-(T/T_c)^n)^d$ where $d$ is some positive number
and $n$ an integer so that the vector mass goes to zero at $T=T_c$
as has been done by workers in the field overestimates the drop
caused by temperature since the vector mass must stay more less
unchanged until the temperature reaches the flash temperature
$\sim 125$ MeV. Thirdly the fusing will have a compensating
effect, e.g., the quantum mechanical level repulsion, between the
``sobar" configuration and the ``elementary" mode, both subject to
the constraints by vector manifestation, which could shift the
peak away from the naive tree-level pole position.


\end{document}